 \documentclass[final,5p,times,twocolumn,number]{elsarticle}

\usepackage{amssymb}
\usepackage{lipsum}
\usepackage{amsmath}
\usepackage{dblfloatfix}
\usepackage{comment}
\biboptions{sort&compress}

\journal{Physics Letters B}

\begin{document}

\begin{frontmatter}



\title{Self-field of a moving string}

\author{Richard A. Battye}
\ead{Richard.Battye@manchester.ac.uk}
\author{Lukasz P. Bunio}
\ead{lukasz.bunio@postgrad.manchester.ac.uk}
\author{Steven J. Cotterill}
\ead{steven.cotterill@manchester.ac.uk}
\affiliation{Jodrell Bank Centre for Astrophysics, Department of Physics and Astronomy, University of Manchester, Oxford Road, Manchester M13 9PL, United Kingdom}

\begin{abstract}

We discuss the effectiveness of the recently proposed quantity $J_{s}$ in suppressing the contribution of a string's self-field to its spectrum, which is used in field theory simulations to track the emission of axions by decaying cosmic strings. We compute the contribution of the self-field to the spectrum of this quantity and to the usually computed spectrum of $\phi\partial_{t}\alpha$ for an infinitely long straight string. Although we demonstrate that the $J_s$ approach is a substantial improvement, we also point out that this highly symmetric model doesn't capture the full contribution of the self-field to the spectrum emitted by strings of a more complex shape, which are typical in network simulations. We then illustrate this point numerically using a simulation of a sinusoidally perturbed straight string. For this simple configuration, we managed to separate the contribution from the self-field and the radiation to the spectrum of $J_{s}$, using the self-field subtraction method. This allows us to show that the spectrum of $J_{s}$ is still dominated by the $n=1$ mode of the string's oscillation, which can be attributed to the self-field and is largely suppressed when the self-field is removed. We also demonstrate that this mode is predominantly sourced by variations along the direction parallel to the string, which is missing in the unconnected segment model, used by \cite{correia2025} to claim the effectiveness of $J_{s}$ in suppressing the self-field's contribution to the spectrum. 

\end{abstract}

\end{frontmatter}

\section{Introduction}
\label{introduction}
Global cosmic strings are topological defects that are produced by the Kibble-Żurek mechanism \cite{TWBKibble_1976,ZUREK1996177} in many extensions of the Standard Model of particle physics that feature a breaking of a global U(1) symmetry in the early universe. One of the most studied models featuring global cosmic strings is the QCD axion model introduced by Peccei and Quinn \cite{peccei1977constraints} and incorporated into the framework of the Standard Model, e.g. in DFSZ \cite{ref:DFSZ, ref:Zhit} and KSVZ \cite{ref:K, ref:SVZ} models. Thanks to its properties, the axion can be a good dark matter candidate \cite{marsh2016axion}, which is produced predominantly by the decay of the axion strings after the Peccei-Quinn symmetry breaking or by the Initial Misalignment Mechanism \cite{ref:misalign1, ref:misalign2, ref:misalign3}, depending on the choice of the parameters of the model. Understanding the production of axions by cosmic strings is, therefore, of great importance; however, it is challenging due to the non-linear nature of the defects' dynamics. Hence, the production mechanism is usually studied numerically, e.g. in field theory simulations, where a network of strings is formed and decays over time, emitting axions \cite{Kim:2024wku, Benabou:2024msj, buschmann2020early, buschmann2022dark,Saikawa:2024bta,Kaltschmidt:2025nkz,PhysRevD.83.123531,Gorghetto:2020qws,Gorghetto:2018myk,correia2025}.

In this letter, we will concentrate on the details of extracting the axion spectrum from numerical simulations, which is used to calculate the instantaneous emission spectrum that contains information about the distribution of energies carried by the axions and their number density. This density is then compared against the energy density of dark matter measured by cosmological probes, e.g. Planck \cite{Planck:2018vyg} , to infer an estimate of the axion mass. It is, however, a highly non-trivial task as noted in \cite{Battye:2026whd}, as the motion of strings with respect to the simulation frame may contribute to the spectrum calculated based on the simulation data, affecting the prediction of the axion mass. Hence, estimating and removing the contribution of the self-field of strings to the spectrum is an essential step in verifying the accuracy of the prediction. 

In section \ref{sec_l_str}, we will define the spectra of quantities used to track the radiation emitted by the strings; namely $\phi\partial_t{\alpha}$ and the recently introduced $J_{s}$ \cite{correia2025},  and we break them down into the contributions from the self-field and axion radiation. We will then concentrate on the spectrum of the self-field’s contribution and present an analytic calculation of its spectrum for an infinitely long straight string moving at a constant velocity $v$. Due to its simplicity, this model can be used to assess the fraction of the spectrum that is sourced by the self-field, as was done in \cite{correia2025}, using an unconnected segment model \cite{PhysRevD.55.573,PhysRevD.59.023508,PhysRevD.60.083504,PhysRevD.86.123513,PhysRevD.93.123503}. In the next section, we will briefly summarise a numerical approach used in \cite{Battye:2026whd} to calculate the spectrum of axions emitted by the string, in which a self-field is subtracted following the self-field subtraction methodology developed in \cite{Davis:1989nj, battye1994global, Battye:2026whd}. We will apply this approach to the quantity $J_{s}$, and show a substantial difference between spectra of subtracted and non-subtracted $J_{s}$. Nevertheless, we will also see that compared to the more commonly used spectrum of $\phi\partial_{t}\alpha$, $J_{s}$ performs much better at suppressing the self-field. In section \ref{sec_com} we comment on the source of the difference observed in a previous section and point out why the model described in section \ref{sec_l_str} and used in \cite{correia2025} does not capture all features of the spectrum of self-field and may underestimate it’s influence on the spectrum of $J_{s}$ calculated for the random network of strings.

\section{Self-field of the infinitely long moving string}
\label{sec_l_str} 

The motion of the string can be seen as a motion of the observer relative to the frame in which the string is static. The coordinates of the string and observer's reference frames are related by the Lorentz transformation of the coordinate system in the direction of motion of the string, and are given by 
\begin{equation}
\label{eq_1}
        t' = \gamma(t-vx), \quad x' = \gamma(x-vt), \quad y' = y, \quad z' = z  ,  
\end{equation}
where the coordinates of string's rest frame are given by primed variables and unprimed coordinates are the simulation frame in which the string is moving with constant velocity $v$ in the $x$ direction. In this case, a string is not expected to radiate as there is no acceleration acting on it, and hence the entirety of the spectrum can be attributed to the self-field. In the string's reference frame, the string is described by the Abrikosov-Nielsen-Olesen vortex solution~\cite{Abrikosov:1956sx, Nielsen:1973cs} and possesses a cylindrical symmetry, thanks to which the solution can be written as 
\begin{equation}
\label{eq_2}
    \Phi(x',y',z') = \phi(r')e^{i\theta'},
\end{equation}
where $r' = \sqrt{x'^2+y'^2}$ and $\theta'$ is an angular coordinate in x'-y' plane, that is given by $\theta' = \arctan\left(\frac{y'}{x'}\right)$. Due to the system's translational symmetry in the z direction, the solution is independent from z, which allows us to treat the problem as effectively two dimensional. Using the transformation equations given above and the expression for $\theta'$, we can write the following derivatives with respect to the simulation frame
\begin{equation}
\label{eq_3}
    \begin{gathered}
        \partial_{x}\theta' = -{\gamma y\over \gamma^2(x-vt)^2+y^2}\approx -{\sin\theta\over r}\left[1+{2vt\cos\theta\over r}+{\cal O}(v^2)\right],\\
        \partial_{y}\theta' = {\gamma (x-vt)\over \gamma^2(x-vt)^2+y^2}\approx{\cos\theta\over r}\left[1+{2vt\cos\theta\over r}\right]-\frac{vt}{r^2}+{\cal O}(v^2),\\
        \partial_{t}\theta' = {\partial\theta^\prime\over\partial x^\prime}{\partial x^\prime\over\partial t}={v\gamma y\over \gamma^2(x-vt)^2+y^2} = -v\partial_{x}\theta',\\
    \end{gathered}
\end{equation}
where we expanded the expressions to first order in velocity in the limit $v \rightarrow0$. We note that $\partial_{z}\theta' = 0$, which will be important for the argument made in the subsequent sections. In the small velocity limit, we can also write
\begin{equation}
\label{eq_4}
    \phi(r') \approx\phi(r)-vt \cos(\theta)\frac{d\phi}{d r'}\bigg\lvert_{r' = r}+{\cal O}(v^2).
\end{equation}
Using these expressions we can now compute the contribution of the moving string to the spectrum of $\phi\partial_{t}\alpha$, that is usually computed in the network simulations \cite{Kim:2024wku, Benabou:2024msj, buschmann2020early, buschmann2022dark,Saikawa:2024bta,Kaltschmidt:2025nkz,PhysRevD.83.123531,Gorghetto:2020qws,Gorghetto:2018myk,correia2025}, and that was investigated in detail in \cite{Battye:2026whd}. We will also compute a contribution of the spectrum of self-field  to the alternative quantity $J_{s}$ that was recently proposed in \cite{correia2025}. In what follows, we will compare the two quantities, attempting to assess which of them is a cleaner observable, for the axion radiation emitted by the strings.

\subsection{$\phi\partial_{t}\alpha$ }
The spectrum of $\partial_{t}\alpha$ can be defined as
\begin{equation}
\label{eq_5}
     \frac{\partial \rho_{\alpha}}{\partial k} = \frac{k^2}{(2\pi L)^3}\int \lvert \widetilde{\phi\partial_t\alpha} \rvert^2d\Omega_{\rm k}\,,
\end{equation} 
 where $L$ is the size of the box, $\Omega_{\rm{k}}$ is the solid angle in momentum space, and we will denote the Fourier transform of $a$ with $\Tilde{a}$. Within the methodology of self-field subtraction developed in \cite{Davis:1989nj, battye1994global, Battye:2026whd}, we can separate  the phase of the field as $\alpha = \theta +\Delta \alpha$, where $\theta$ is the background field, sourced by the string, and $\Delta\alpha$ is a small perturbation around it that can be identified as radiation emitted by the string. Using these fields, we can express the spectrum as a sum of three terms
\begin{equation}
\label{eq_6}
     \frac{\partial \rho_{\alpha}}{\partial k} = \int  \frac{k^2d\Omega_{\rm k}}{(2\pi L)^3}\left\{\lvert\widetilde{\phi\partial_{t}\theta}\rvert^2 +2Re\left[\widetilde{\phi\partial_{t}\theta}(\widetilde{\phi\partial_{t}(\Delta \alpha)})^* \right] +\lvert\widetilde{\phi\partial_{t}(\Delta \alpha)}\rvert^2 \right\}\,,
\end{equation} 
where the first term is the spectrum of self-field (that we will call $\frac{\partial \rho_{\theta}}{\partial k}$) the last term is the spectrum of radiation $\left(\frac{\partial \rho_{\Delta\alpha}}{\partial k}\right)$, and the term in the middle is a cross term. For a string moving at a constant velocity, only the first of these terms is non-zero and can be calculated analytically. We start by defining the Fourier transform of $\phi\partial_{t}\theta$ that can be written as 
\begin{equation}
    \label{eq_7}
    \widetilde{\phi\partial_{t}\theta} = v\gamma \int e^{-ik_{z}z}dz \iint dx dy \phi(r'){ y\over \gamma^2(x-vt)^2+y^2} e^{-i(k_{x}x+k_{y}y)}\,,
\end{equation}
where we used expression (\ref{eq_3}).
We can now do the integral over z and change the coordinate system of the remaining two integrals to polar coordinates
\begin{equation}
    \label{eq_8}
    \widetilde{\phi\partial_{t}\theta} = 2\pi v \delta(k_{z}) e^{-ik_{\perp}vt\cos\varphi_k}\int d\rho\phi(\rho) \int d\beta \sin\beta e^{-ik_{\perp}\rho\kappa \cos(\beta-\beta_{0})}\,,
\end{equation}
where $k_{\perp} = k\sin\theta_{k}$ is the magnitude of the component of momentum perpendicular to the string, while the factor of $\delta(k_z)$ represents the translational invariance along the string. To get this expression we first substituted $X = \gamma(x-vt)$ and then defined a polar coordinate system such that $\rho = \sqrt{X^2+y^2}$ in which $\beta$ is the polar angle, as well as the definitions $\tan\beta_{0} = \gamma\tan\varphi_{k}$, $\kappa = \sqrt{\sin^2\varphi_{k}+\frac{1}{\gamma^2}\cos^2\varphi_{k}}$, where $\varphi_{k}$ is the polar angle in the momentum space conjugate to $x-y$ coordinates. Since $\phi$ is only a function of the radial coordinate, the second integral in this expression can be evaluated analytically and gives
\begin{equation}
\label{eq_9}
    \widetilde{\phi\partial_{t}\theta} = -(2\pi)^2 iv \delta(k_{z}) e^{-ik_{\perp}vt\cos\varphi_k}\sin\beta_{0}\int d\rho\phi(\rho) J_{1}(\kappa k_{\perp} \rho)\,,
\end{equation}
where $J_{1}(\kappa k_{\perp} \rho)$ is a Bessel function of the first kind. The remaining integral over $\rho$, that we will from now on call $I$, can be performed numerically in a small velocity limit in which $\kappa\rightarrow1$, and $J_{1}(\kappa k_{\perp} \rho) = J_{1}(k_{\perp} \rho)+{\cal O}(v^2)$. Hence the integral $I$ is independent of velocity up to second order in $v$, and one can clearly see that $\widehat{\phi\partial_{t}\theta}$ scales as $v$, as was shown in \cite{Battye:2026whd} and \cite{correia2025}. The final step is to compute the spectrum by integrating the square of the modulus of expression (\ref{eq_9}) over the solid angle (again in the small velocity limit, in which I and $\kappa$ are independent from $\varphi_{k}$). As a result, we get
\begin{equation}
\label{eq_10}
\frac{\partial \rho_{\theta}}{\partial k} = \frac{k\pi v^2 I^2}{L^2}\,, 
\end{equation}
where we used the properties of the $\delta$-function to write $\delta(k_{z}) = \frac{\delta(\theta_{k} - \frac{\pi}{2})}{k\sin\theta_{k}}$. After performing the integral over $\theta_{k}$, the variable $k_{\perp} = k$ and one of the $\delta$-functions remains in the expression as $\delta(0)$. We assign meaning to it using a finite volume regularization scheme in which $\delta(0) = \frac{L}{2\pi}$.

\subsection{$J_{s}$ }

The other quantity we want to investigate, $J_{s}$, was defined in \cite{correia2025}  as
\begin{equation}
\label{eq_11}
    J_s(k)={\hat k}_i\Tilde{J}_i(k_x,k_y, k_z)\,,
\end{equation}
where $\hat{k}_{i}$ is the unit vector in momentum space, $J_{i} = \phi(r)\partial_{i}\alpha$ and index $i$ runs over $x,y,z$. The quantity is constructed in a similar way to the diagnostic introduced in \cite{PhysRevD.105.063517, PhysRevD.107.043507}, which is obtained by taking a radial projection of the vector $J_{i}$ in coordinate space and has been shown to be highly effective in suppressing the self-field \cite{PhysRevD.105.063517, PhysRevD.107.043507, Battye:2026whd}. The spectrum of $J_{s}$, which we will call $\frac{\partial J_{s}}{\partial k} $, can be defined analogously to spectrum $\frac{\partial \rho_{\alpha}}{\partial k} $ as
\begin{equation}
\label{eq_12}
     \frac{\partial J_{s}}{\partial k}= \frac{k^2}{(2\pi L)^3}\int \lvert J_{s}(\mathbf{k}) \rvert^2 d\Omega_{\rm k}\,.
\end{equation}
This quantity also admits a decomposition similar to the one we performed for $\phi\partial_{t}\alpha$, given by
\begin{equation}
\label{eq_13}
     \frac{\partial J_{s}}{\partial k}= \int \frac{k^2d\Omega_{\rm k}}{(2\pi L)^3}\left\{\lvert J_{\theta}(\mathbf{k}) \rvert^2+2Re[ J_{\theta}(\mathbf{k})(J_{\Delta\alpha}(\mathbf{k}))^*]+\lvert J_{\Delta\alpha}(\mathbf{k}) \rvert^2\right\} \,,
\end{equation}
where
\begin{equation}
\label{eq_14}
     J_{\theta}(\mathbf{k}) = \hat{k}_{i}\widetilde{\phi\partial_{i}\theta}, \quad
     J_{\Delta\alpha}(\mathbf{k}) = \hat{k}_{i}\widetilde{\phi\partial_{i}\Delta\alpha}\,,
\end{equation}
and we can recognize the same structure of respectively self-field $\left(\frac{\partial J_{\theta}}{\partial_{k}}\right)$, cross-term and radiation $\left(\frac{\partial J_{\Delta\alpha}}{\partial_{k}}\right)$ contributions as in previously considered case. 

We proceed by computing the spectrum of the self-field contribution to this quantity. Using expressions (\ref{eq_3}), we can write $J_{\theta}$ as
\begin{equation}
\label{eq_15}
\begin{gathered}
        J_{\theta}=2\pi\delta(k_{z})\gamma\int dxdy\,\phi(r')\left[-y{\cos\varphi_k+(x-vt)\sin\varphi_k}\over \gamma^2(x-vt)^2+y^2\right]\\\times e^{-ik_{\perp}r\cos(\varphi-\varphi_k)}\,,
\end{gathered}
\end{equation}
using the substitutions $X, \rho, \kappa$ and $\beta$ defined in previous subsection we can rewrite it as 
\begin{equation}
\label{eq_16}
    \begin{gathered}
        J_{\theta}=2\pi\delta(k_{z}) e^{-ik_{\perp}vt\cos\varphi_k}\int d\rho d\beta\left[-\sin\beta\cos\varphi_k+{\cos\beta\over\gamma}\sin\varphi_k \right]\\\times\phi(\rho)e^{-ik_{\perp}\rho\left({\cos\beta\over\gamma}\cos\varphi_k+\sin\beta\sin\varphi_k\right)}\,.
    \end{gathered}
\end{equation}
The integral over the angle $\beta$ can be done analytically, giving as a result
\begin{equation}
\label{eq_17}
\begin{gathered}
J_{\theta}={(2\pi)^2\delta(k_{z}) iv^2 \sin(2\varphi_k) e^{-ik_{\perp}vt\cos\varphi_k}\over2\kappa}\int_0^\infty d\rho\,\phi(\rho)J_1(\kappa k_{\perp}\rho)\,,
\end{gathered}
\end{equation}
and hence, in the small velocity limit, the spectrum of self-field can be expressed as
\begin{equation}
\label{eq_18}
\frac{\partial J_{\theta}}{\partial k}= \frac{\pi kv^4I^2}{4L^2}\,.\end{equation}

We can clearly see, comparing expression (\ref{eq_10}) with (\ref{eq_18}), that the self-field contribution of $J_s$ is suppressed by a factor $\propto v^2$ with respect to (\ref{eq_10}), as reported in \cite{correia2025}. Hence, the spectrum scales in proportion to $v^4$. We see that, for a straight string $J_s$ is a quantity that tracks the radiation component much better compared to the usually computed $\phi\partial_{t}\alpha$. However, the model is based on the assumption of cylindrical symmetry which forces $\phi\partial_{z}\theta = 0$. Thanks to this condition, the factor given in the square bracket of equation (\ref{eq_16}) cancels up to second order in $v$, where $\gamma \approx 1$ and $\beta \approx \varphi_k$, which wouldn't necessarily happen if $\phi\partial_{z}\theta \neq 0$. In the case of a general string configuration, the velocity with which different parts of the string move varies along the string, giving rise to non-zero contribution of $\phi\partial_{z}\theta$ to $J_{s}$.One also expects strings to be curved, which would likely generate further contribution to $J_{s}$, as it also breaks the cylindrical symmetry. The size of this contribution is difficult to estimate analytically, but we will illustrate this point numerically in section \ref{sec_com}.

\subsection{Numerical evaluation of the integral I}
\begin{figure*}[!b]
	\centering \includegraphics[width=1\textwidth]{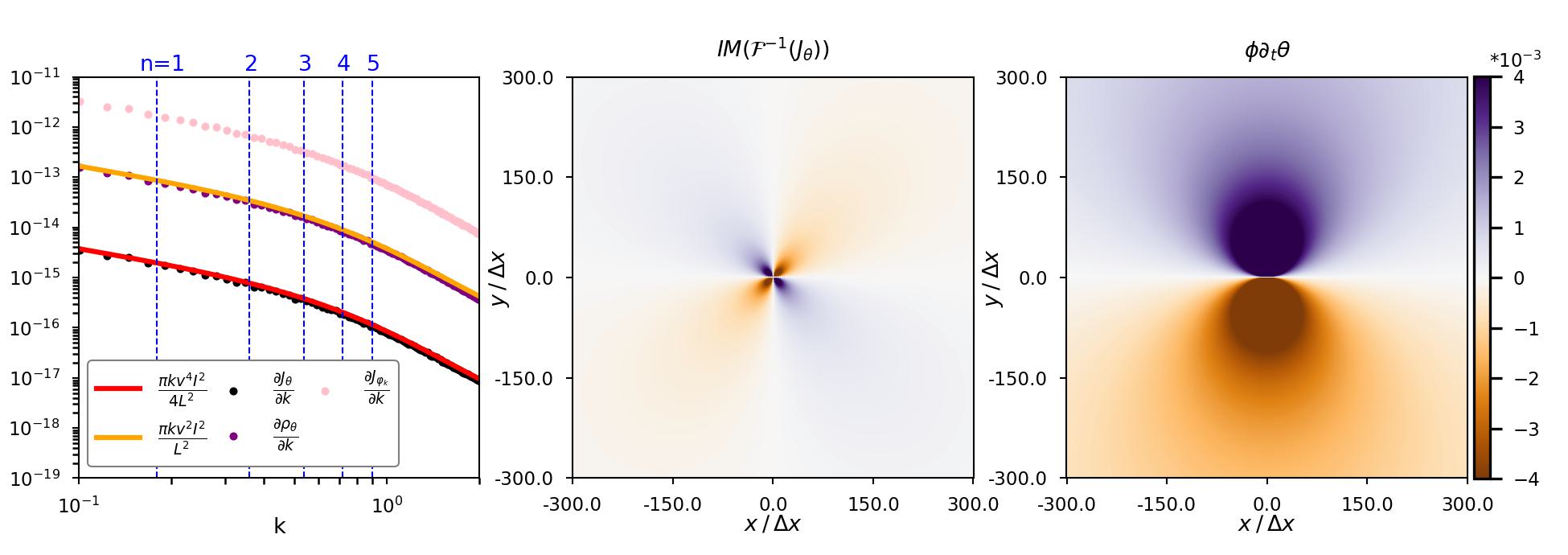}	
	\caption{The field $\phi\partial_{t}\theta$ (right-hand panel) and the imaginary part of the inverse Fourier transformed $J_{\theta}$ (middle panel) for a straight string moving at $v= 0.3$, calculated numerically. Spectra of these fields are shown in the panel on the left with black and purple dots. The red and orange lines show the analytic predictions for the spectra of both of these quantities as given by equations (\ref{eq_10}) and (\ref{eq_18}) times an additional factor $\pi^2/n_{x}^3$ that accounts for the difference in normalisations of discrete and continuous Fourier transforms. The pink dots show a spectrum of the $\hat{\boldsymbol{\varphi}}_{k}$ projection of the Fourier transform of $J_{i}$, while the spectrum of the projection on $\hat{\boldsymbol{\theta}}_{k}$ direction is zero. The vertical dashed lines show the positions of the first 5 peaks if a string had been perturbed according to equation (\ref{eq_23}). There is very clear agreement between the analytic and numerical calculations} 
	\label{fig_1}%
\end{figure*}
Both spectra of the self-field in the small velocity limit depend only on the integral I that is given by
\begin{equation}
\label{eq_19}
    I = \int_{0}^{\infty}d\rho \phi(\rho)J_{1}(k\rho)\,.
\end{equation}
The integral can be evaluated numerically by first solving for the profile function, $\phi(\rho)$. We then note that $\lim_{\rho\rightarrow\infty}\phi(\rho) = 1$ (since we set $\lambda=\eta=1$ in equation (\ref{eq_22})), which means that within numerical precision of the solution there exists a value $\rho_{ct}$ such that for $\rho>\rho_{ct}$ the value of the profile function is equal to 1. Therefore, we can write 
\begin{equation}
\label{eq_20}
    I = \int_{0}^{\rho_{ct}}d\rho \phi(\rho)J_{1}(k\rho) + \int_{\rho_{ct}}^{\infty}d\rho J_{1}(k\rho)\,.
\end{equation}
The second integral can be done analytically using the properties of the Bessel functions, while we compute the first integral numerically. To do that, we discretise $\rho$ variable and use a rectangle rule for integration, which gives
\begin{equation}
\label{eq_21}
    I(k) = \sum_{a = 0}^{a = \rho_{ct}/\Delta\rho}\Delta\rho \phi(a\Delta\rho)J_{1}(ka\Delta\rho) + \frac{J_{0}(k\rho_{ct})}{k},
\end{equation}
that is evaluated for the central value of each $k$-bin.

\section{Numerical simulation}
\label{sec_num}
\begin{figure*}[!b]
	\centering \includegraphics[width=1\textwidth]{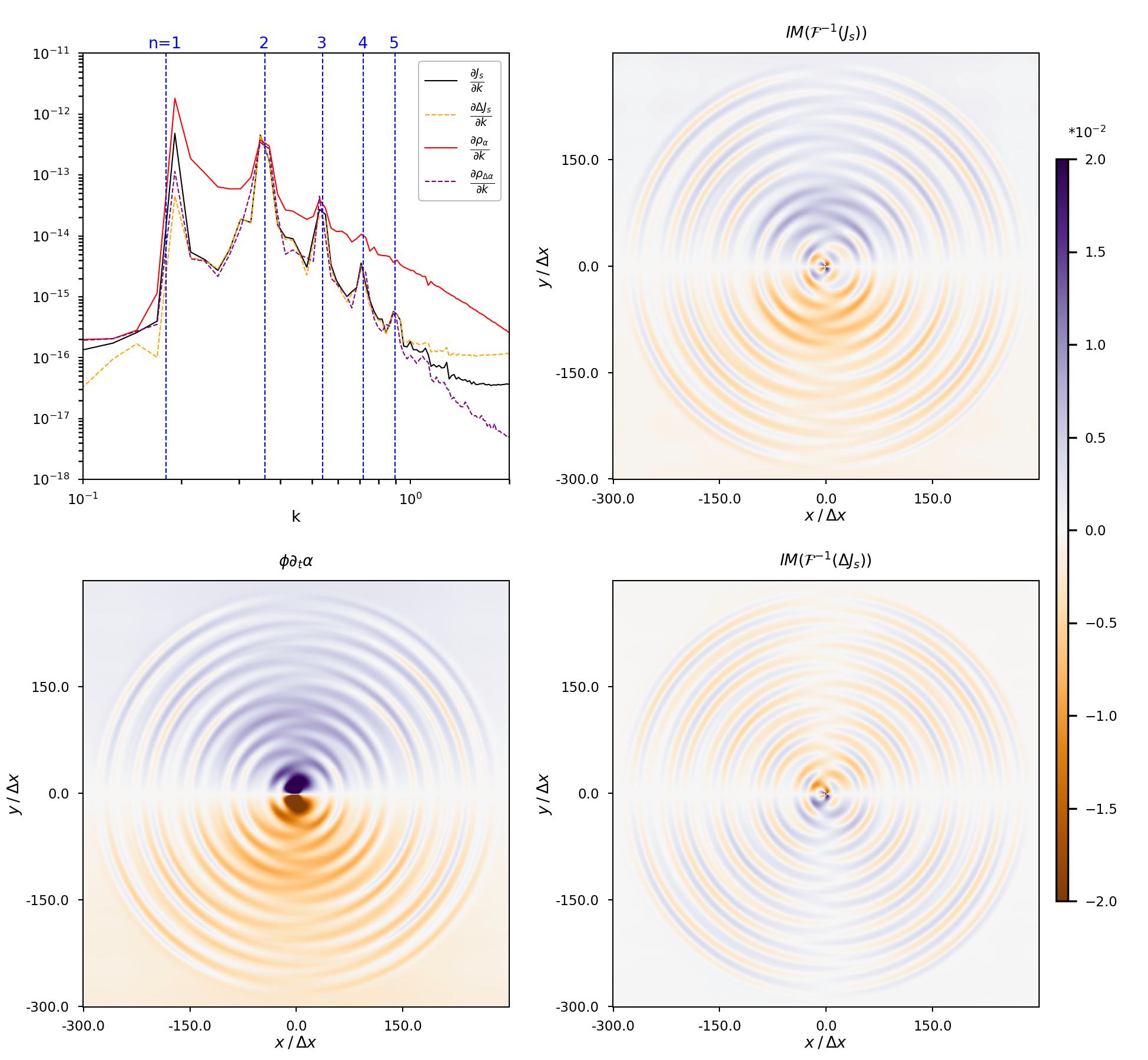}	
	\caption{The heat maps show the $z = 412\Delta x$ slice through the simulation box of the field $\phi\partial_{t}\alpha$ (bottom left panel), and the imaginary part of the inverse Fourier transformed $J_{s}$ (top right panel) and its self-field subtracted counterpart (bottom right panel) for a sinusoidally oscillating string at $t = 678\Delta t$, when a string is approximately straight, initialized with $\varepsilon_{0} = 0.5$. Spectra of these fields are shown in the panel on the top left, together with the positions of the first 5 harmonics shown by vertical dashed lines and the spectrum of $\partial_{t}(\Delta\alpha)$ computed in \cite{Battye:2026whd}.} 
	\label{fig_2}%
\end{figure*}
In this section, we will explain the numerical method used to reanalyse the data produced in \cite{Battye:2026whd}, which will form a basis for the comparison presented in the next section. We will start by briefly summarising the simulation set-up (for details see \cite{Battye:2026whd}). 

The field equations used by the simulation can be obtained using a U(1) symmetric Lagrangian density given by
\begin{equation}
\label{eq_22}
{\cal L}={1\over 2}|\partial_\mu\Phi|^2-{1\over 4}\lambda\left(|\Phi|^2-\eta^2\right)^2\,,
\end{equation}
where $\lambda$ is a coupling constant and $\eta$ is a symmetry-breaking scale, which, without loss of generality, can be both set to 1. The simulation uses a discretised version of this equation to evolve the field on a 3 dimensional grid with $n_{x} = 801$ points in each spatial direction and spacing $\Delta x = 0.7$ between the points. The temporal direction is also discretised, and the simulations ran for $\approx1000$ time steps, that is until the light crossing time, with time steps separated by $\Delta t=0.3$. At the boundaries, the simulation used absorbing boundary conditions, developed in \cite{battye1994global} in the $x-$ and $y-$ direction and a fixed boundary condition in the z-direction. 

The field is initialised as static, using the vertex solution (\ref{eq_2}) for each z-slice offsetting the beginning of the coordinate frame according to
\begin{equation}
\label{eq_23}
    x_{0} = \frac{\varepsilon_{0}L}{2\pi}\sin{\left(\frac{2\pi z}{L}\right)}\,,
\end{equation}
where $\varepsilon_{0} = 0.5$ and the oscillation wavelength $L = 50 \Delta x$. The value of the modulus of the field was assigned using a profile function $\phi_{\rm sol}$, which was obtained numerically.
\begin{figure*}[!b]
	\centering \includegraphics[width=1\textwidth]{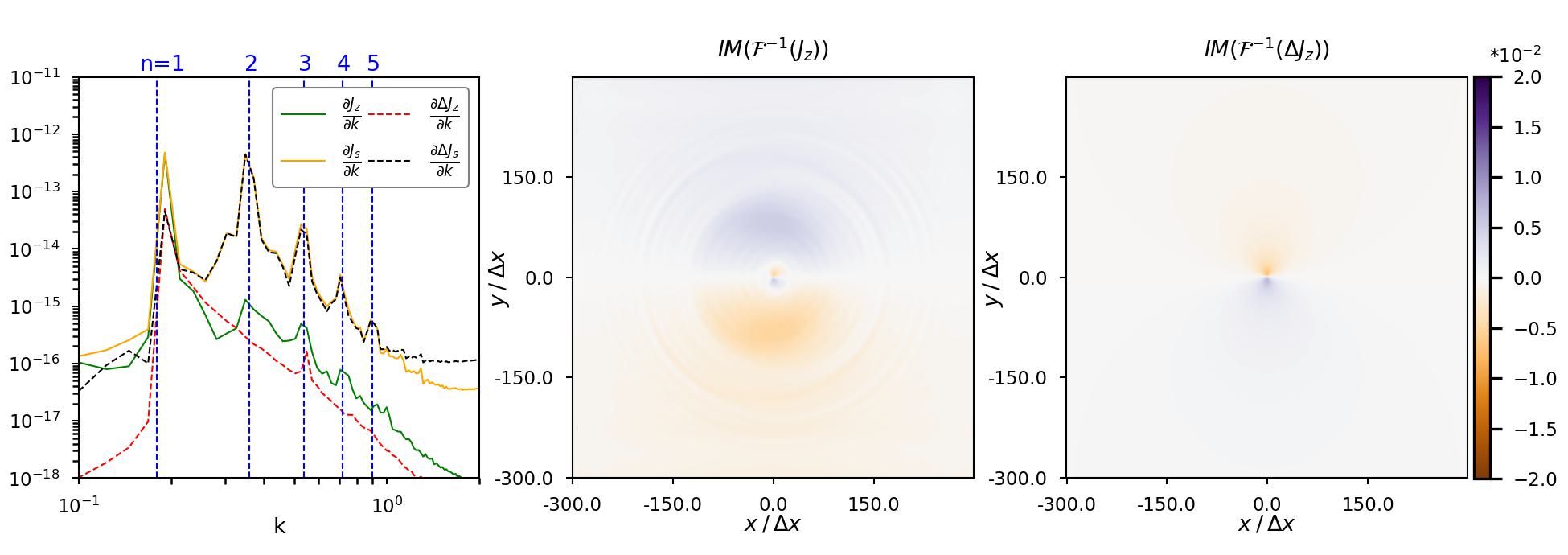}	
	\caption{The heat maps show the $z = 412\Delta x$ slice through the simulation box of the imaginary part of the inverse Fourier transformed $J_{z}$ (middle) and its self-field subtracted counterpart (right) for a sinusoidally oscillating string at $t = 678\Delta t$ initialized with $\varepsilon_{0} = 0.5$. Spectra of these fields together with the spectra of $J_{s}$ and $\Delta J_{s}$ are shown in the panel on the left. The blue dashed vertical lines show the positions of the first 5 harmonics.} 
	\label{fig_3}%
\end{figure*}
Using this simulation data, we want to separate the contribution due to radiation from the self-field. We can do this by using the following approximation 
\begin{equation}
    \label{eq_24}
    \phi\partial_\mu(\Delta\alpha)\approx\phi\partial_\mu\alpha-\phi_{\rm sol}\partial_\mu\theta\,,
\end{equation}
where $\phi,\alpha$ are the modulus and phase of the field output by the simulation, $\theta$ is an angular variable of the polar coordinate system in the $x-y$ plane centred at the position of the string and index $\mu$ runs over $\{t, x,y,z\}$. The derivatives of the self-field's phase are computed using equations (\ref{eq_3}) at $t=0$, and the derivative with respect to $z$ is computed numerically. The velocity of a string is computed using a first-order numerical derivative of the string's positions, which are detected by the algorithm described in \cite{Battye:2026whd}. The quantity $\phi\partial_\mu(\Delta\alpha)$ is used to construct $J_{s}$ and to compute the following spectra 
\begin{equation}
\label{eq_25}
\begin{gathered}
     \frac{\partial \rho_{\Delta\alpha}}{\partial k} = \frac{k^2}{(2\pi L)^3}\int  \lvert\widetilde{\phi\partial_{t}(\Delta \alpha)} \rvert^2 d\Omega_{\rm k}\,,\\
     \frac{\partial J_{\Delta\alpha}}{\partial k}= \frac{k^2}{(2\pi L)^3}\int \lvert J_{\Delta\alpha} \rvert^2 d\Omega_{\rm k}.
     \end{gathered}
\end{equation}
To assess the impact of the velocity's variation along the string, we will also compute the spectra of the self-field that were computed using the field $\phi_{\rm{sol}}\partial_{\mu}\theta$ and spectra of $J_{z}$ and its self-field  subtracted counterpart for which the sum over index $i$ in the definition of $J_{s}$ \ref{eq_11} runs only over $z$.

The numerical Fourier transforms are performed using the FFTW3 library \cite{FFTW05}. The solid angle integral is discretized using the following prescription
\begin{equation}
\label{eq_26}
    \frac{k^2}{(2\pi L)^3}\int d\Omega_{k}\rightarrow\frac{(dk)^3}{(2\pi L)^3\Delta k}\sum_{k = k'-\Delta k/2}^{k = k'+\Delta k/2}
\end{equation}
where $k'$ is the central value of the k-bin, $dk$ is the distance between two neighbouring points of the lattice in momentum space, and $\Delta k$ is the width of the momentum bin given by $\frac{2\pi}{\Delta x nx} = dk$. More details about this calculation can be found in the appendix of \cite{Battye:2026whd}.

\section{Comparison of the methods}
\label{sec_com}

We start by comparing the spectra calculated for an infinitely long straight string. The left-hand panel of figure \ref{fig_1} shows spectra (\ref{eq_10}) and (\ref{eq_18}), which we calculated in section \ref{sec_l_str}, for a string moving at $v = 0.3$. To verify our numerical approach, we also calculated the same quantities numerically for a straight-string moving in the $x$ direction with the same velocity. To do that, we used a Lorentz-boosted straight string solution as initial conditions and periodic boundary conditions in the direction perpendicular to the string. We see that the numerical results follow the analytic prediction very well. The panel also shows the spectrum of the quantity formed by projecting the vector $\Tilde{J_{i}}$ on the $\hat{\boldsymbol{\varphi}}_{k}$ direction, while the spectrum of the projection $\Tilde{J_{i}}$ on the $\hat{\boldsymbol{\theta}}_{k}$  direction is up to numerical precision equal to zero. The spectrum of the $\hat{\boldsymbol{\varphi}}_{k}$ projection follows the same shape as that of $J_{\theta}$, but its values are larger by three orders of magnitude, and as noted in \cite{correia2025}, it would dominate the energy spectrum of a moving string. The middle panel shows the numerically computed inverse Fourier transform of $J_{\theta}$. The field follows a quadrupolar shape, which can be predicted by inverse transforming expression (\ref{eq_17}), and is visibly fainter than the field shown on the right. This panel shows the self-field's contribution $\phi\partial_{t}\theta$, which in turn follows a dipolar pattern, as predicted by equation (\ref{eq_9}) and in \cite{Battye:2026whd}. Comparing the spectra shown on the left, and the fields shown in the middle and on the right, one might conclude that the contribution of the self-field is strongly suppressed in $J_{s}$ compared to the traditional $\phi\partial_{t}\alpha$. This is certainly true for the case of a straight string in linear motion, which is a highly symmetric configuration. However, as we will now show, there remains a sizeable contribution from the self-field when this symmetry is removed.

As an example of this, we now turn to a simulation of a sinusoidally perturbed string, which we consider to be a relatively mild departure from the above. The results are shown in figure \ref{fig_2}, where we plot the inverse Fourier transforms of $J_{s}$ (top right panel), its self-field subtracted counterpart (bottom right panel) and $\phi\partial_{t}\alpha$ (top left panel). As reported in \cite{Battye:2026whd}, the $\phi\partial_{t}\alpha$ features a strong dipolar pattern in the centre, with small waves propagating radially away from the string on top of it, which can be identified as axion radiation. In comparison, the inverse Fourier transform of $J_{s}$ is much more dominated by the contribution from the radiation than $\phi\partial_{t}\alpha$; however, a small dipole contribution is still visible beneath it. The dipole is almost completely removed only in the inverse Fourier transform of $\Delta J_{s}$, where we can see that the field is dominated by the quadrupolar structure of the radiation. These features are quantified in the spectra of these quantities shown in the left-hand panel at the top. There we can see that the spectrum of $\rho_{\alpha}$ is dominated by the $n=1$ peak. The peaks that, based of the analytic argument presented in \cite{battye1994global}, can be associated with the radiation: $n = 2,3,4,5 ...$, are subdominant, and higher $n$ peaks are hidden below a large background of the self-field similar to that shown in figure \ref{fig_1}. The continuous background is highly suppressed in the case of the spectra of $J_{s}$ and $\Delta J_{s}$. However, the $n=1$ peak in the spectrum of $J_{s}$ is lower by a factor of $\approx0.26$ than in the spectrum of $\rho_{\alpha}$, but it is still at a similar level as the $n=2$ peak. The value of this factor is approximately equal to the value of the root mean square velocity which at this time step is $\approx0.24$. We, therefore, see that the dominant contribution of the self-field to the spectrum, which comes from the $n=1$ peak, is not in general suppressed by the factor of $v^2$ as was claimed in \cite{correia2025} based on the independent segment approximation --- very similar in nature to the model presented in section \ref{sec_l_str}. Even though the height of the $n=1$ peak is reduced in the spectrum of $J_{s}$, compared to spectrum of $\phi\partial_{t}\alpha$, only after self-field subtraction is performed on $J_{s}$ does the contribution of the $n=1$ peak become subdominant, as it is supressed by an additional factor of $\approx 0.09$. At the same time, this subtraction doesn't change the height of the other peaks.

To understand why that is the case, we separated out the $J_{z} = \phi\partial_{z}\alpha$ component of the $J_{s}$ quantity that is presented in figure \ref{fig_3}. The heat maps show the inverse Fourier transform of $J_{z}$ in the middle, and its self-field subtracted version is shown on the right. We can immediately recognize the faint background with the dipolar structure that we noticed in the heat map of $J_{s}$ shown in figure \ref{fig_2}. The self-field subtraction visibly suppresses this contribution to the field, as can be  seen by comparing the panel in the middle and on the right. We can also see this effect in the spectra shown in the panel on the left, where we compare the spectra of $J_{s},\Delta J_{s}, J_{z}$ and $ \Delta J_{z}$. Notably, $J_s \approx J_z$ and $\Delta J_s \approx \Delta J_z$ in the vicinity of the first peak and, hence, we can conclude that this is the dominant contribution to $J_s$ around $n=1$. We also note that $J_z$ is the component of the current in which the breakdown of cylindrical symmetry is manifest.

We can see this if we consider a Lorentz transformation with a velocity that is a function of $z$. Effectively that corresponds to performing the transformation in 2D for each $x-y$ slice separately. $\phi\partial_{x}\theta$ and $\phi\partial_{y}\theta$ remain the same as in the case of a straight string moving at uniform velocity as given by (\ref{eq_3}), however $\phi\partial_{z}\theta$ is no longer vanishing and is given by

\begin{equation}
\label{eq_27}
\phi\partial_{z}\theta = \frac{-y\gamma^3}{\gamma^2(x-vt)^2+y^2}(vx-t)\frac{\partial v}{\partial z}.
\end{equation}
If we performed the calculation with this ansatz instead of uniformly moving string, we would get the same result for $J_{\theta}$ plus a contribution from $J_{z}$ that scales $\propto v\frac{\partial v}{\partial z}$ which isn't captured by the straight string model, and as we showed numerically, this contribution is of non-negligible size.

\section{Summary and conclusions}
\label{sec_sum}

In this letter, we compared the contribution of the self-field of a moving string to the spectra of $\phi\partial_{t}\alpha$ and $J_{s}$. We showed by calculating the spectra of a straight string moving with a constant velocity that the spectrum of $J_{s}$ is, in this case, suppressed by a factor $\propto v^2$ with respect to the spectrum of $\phi\partial_{t}\alpha$ as previously shown in \cite{correia2025} for an unconnected segment model. However, in doing so we identified that this effectiveness of $J_{s}$ at suppressing the spectrum is based on an almost exact cancellation between terms that is forced by the cylindrical symmetry of the string (up to a Lorentz boost). We then verified this claim by reanalysing the simulation data produced for \cite{Battye:2026whd}, where we simulated a sinusoidally perturbed straight string and separated the radiation from self-field using the self-field subtraction method. Applying this method to $J_{s}$, we managed to separate the spectrum of radiation and that of the self-field. We then saw that the spectrum of $J_{s}$ contains a substantially smaller contribution of the self-field than that of the usually computed $\phi\partial_{t}\alpha$; however, it is still dominated by the $n=1$ harmonic of the string's oscillation. By computing the spectrum of the components of $J_{s}$, we were able to demonstrate that this part of the spectrum is sourced by $J_z$, a component that is always zero in the straight string and unconnected segment models.

Based on the spectra shown in figures \ref{fig_2} and \ref{fig_3}, we have shown that, although $J_s$ performs considerably better than $\phi\partial_t\alpha$ at extracting the axion radiation, there is still a significant contribution from the self-field. In \cite{correia2025} the spectrum of $\phi\partial_{t}\alpha$ and $J_{s}$ were computed for a network of strings and displayed in figure 4. There is a striking similarity between the two which leads the authors to claim that there is a relatively small contribution from the self-field. However, it is difficult to reconcile this interpretation with the stark difference between the spectra of $\phi\partial_t\alpha$ and $\phi\partial_t(\Delta\alpha)$ in our relatively controlled scenario. Although it is not trivial to apply our results to the more complicated case of a network, the natural expectation based on what we have presented is that $J_s$ should be less effective at filtering out the self-field in this scenario, due to the highly non-symmetric nature of the configuration. We, therefore, point out that an alternative interpretation of their results is also plausible, namely that the reason their two spectra are so similar is not because the self-field hardly contributes to either measure, but rather because the self-field dominates both measures.

\section*{Acknowledgements}
We are grateful for the discussions with the
participants of the “Universe Unravelled: Cosmic Strings at 50”
workshop held at the Department of Applied Mathematics and Theoretical Physics in Cambridge. We want to thank Mark Hindmarsh, Jose Correia, Amelia Drew and Paul Shellard for their helpful comments and discussion of this work. In addition, we would like to acknowledge the collaboration with Pranav B. G. Manoj on related topics. The work of LB is supported by the STFC Doctoral Training Award No. ST/X508597/1. The simulations presented in this work were conducted on the resources of the Computational Shared Facility at the University of Manchester. 
\appendix
\bibliographystyle{elsarticle-num} 
\bibliography{bibl.bib}
\end{document}